\documentclass[twoside, epsfig]{article}

\input oejv.sty
  \usepackage{amsfonts,amsmath,amssymb,epsfig,lscape}
\def\degr{\hbox{$^\circ$}}
\def\sun{\hbox{$\odot$}}
\def\dmf{\dot{\mathfrak{M}}}
\def\la{\mathrel{\mathchoice {\vcenter{\offinterlineskip\halign{\hfil
$\displaystyle##$\hfil\cr<\cr\sim\cr}}}
{\vcenter{\offinterlineskip\halign{\hfil$\textstyle##$\hfil\cr <\cr\sim\cr}}}
{\vcenter{\offinterlineskip\halign{\hfil$\scriptstyle##$\hfil\cr <\cr\sim\cr}}}
{\vcenter{\offinterlineskip\halign{\hfil$\scriptscriptstyle##$\hfil\cr
<\cr\sim\cr}}}}}
\def\ga{\mathrel{\mathchoice {\vcenter{\offinterlineskip\halign{\hfil
$\displaystyle##$\hfil\cr>\cr\sim\cr}}}
{\vcenter{\offinterlineskip\halign{\hfil$\textstyle##$\hfil\cr
>\cr\sim\cr}}}
{\vcenter{\offinterlineskip\halign{\hfil$\scriptstyle##$\hfil\cr
>\cr\sim\cr}}}
{\vcenter{\offinterlineskip\halign{\hfil$\scriptscriptstyle##$\hfil\cr
>\cr\sim\cr}}}}}
\newcommand{\be}{\begin{equation}}
\newcommand{\ee}{\end{equation}}
\newcommand{\bdm}{\begin{displaymath}}
\newcommand{\edm}{\end{displaymath}}

\begin{document}

 \OEJVhead{August 2008}
 \OEJVtitle{AE~Aquarii: The first white dwarf in the family}
 \OEJVtitle{of spin-powered pulsars}
 \OEJVauth{Ikhsanov, N.R.; Beskrovnaya, N.G.}
 \OEJVinst{Central Astronomical Observatory of the Russian Academy of Sciences at Pulkovo,
 Pulkovskoe Shosse 65-1, St.\,Petersburg 196140, Russia, {\tt ikhsanov@gao.spb.ru}}

\OEJVabstract{Simulation of Doppler H$\alpha$ tomogram of the nova-like star
AE~Aquarii suggests that the dipole magnetic moment of the white dwarf is close to
$1.5 \times 10^{34}\,{\rm G\,cm^3}$. This is consistent with the lower limit to the
magnetic field strength of the white dwarf derived from observations of circularly
polarized optical emission of the system. The rapid braking of the white dwarf and
the nature of pulsing hard X-ray emission recently detected with SUZAKU space
telescope under these conditions can be explained in terms of spin-powered pulsar
mechanism. A question about the origin of strongly magnetized white dwarf in the
system remains, however, open. Possible evolutionary tracks of AE~Aquarii are briefly
discussed.}

\begintext


 \section{Introduction}

Basic parameters of the close binary system AE~Aqr are listed in
Tab.\,\ref{ikhsanov-t1}. The red dwarf overflows its Roche lobe and transfers material
through the L1 point towards the white dwarf. This material, however, is neither
accreted onto the surface of the white dwarf nor stored in a disk around its
magnetosphere. Instead, it is streaming out from the system with an average velocity of
$300\,{\rm km\,s^{-1}}$. The spin-down power of the white dwarf exceeds the bolometric
luminosity of the system by a factor of a few (see Tab.\,\ref{ikhsanov-t2}). Finally,
the system shows flaring activity whose properties are absolutely unique among all
classes of flaring astrophysical objects \citep[for a review see][and references
therein]{Beskrovnaya-etal-1996,Ikhsanov-etal-2004}.

Theoretical studies of the system during the last decade have led to a conclusion that
AE~Aqr does not fit in any of the accretion-based models developed for Cataclysmic
Variables \citep[see, e.g.][]{Wynn-etal-1997}. In other words, the white dwarf in the
system is not an accretion-powered pulsar. This situation can be explained in terms of
the centrifugal inhibition (propeller) model provided that the magnetic field of the
white dwarf is strong enough for its magnetospheric radius to exceed the corotational
radius (see Tab.\,\ref{ikhsanov-t3}). However, how strong is the magnetic field\,? A
lack of the white dwarf's photospheric lines in the system spectra makes impossible a
direct measurement of the field strength through Zeeman effect. Among other methods,
which can be used to answer this question, are the simulation of Doppler H$\alpha$
tomogram of the system, the modeling of the rapid braking of the white dwarf, the
analysis of the circularly polarized optical emission, and a reconstruction of power
source of the pulsing hard X-ray emission recently discovered by SUZAKU space telescope.
Following these methods one comes to a conclusion that the dipole magnetic moment of the
white dwarf in AE~Aqr is close to $1.5 \times 10^{34}\,{\rm G\,cm^3}$ and its rapid
braking is governed by the pulsar-like spin-down mechanism.

\begin{table}[t]
 \caption[]{Parameters of AE~Aquarii$^{*}$}
\label{ikhsanov-t1}
\begin{tabular}{l|ccccc}
  \noalign{\smallskip}
  \hline
  \noalign{\smallskip}
{\it System parameters} &  Distance  & Binary period & Inclination & Eccentricity & Mass ratio  \\
  \noalign{\smallskip}
  \hline
  \noalign{\smallskip}
Value &  $(100\pm30)$\,pc  &  $9.88$\,hr &  $50^{\degr}<i<70^{\degr}$ & $0.02$ & $0.6-0.8$ \\
  \noalign{\smallskip}
  \hline
  \noalign{\smallskip}
  \noalign{\smallskip}
  \hline
  \noalign{\smallskip}
{\it Stellar parameters}   &  Type   & Mass ($M_{\sun}$) & Spin period &  $\dot{P}$ (${\rm s\,s^{-1}}$) & $\left[\vec{\bf \Omega}\,\wedge\,\vec{\bf m}\right]^{**}$\\
  \noalign{\smallskip}
  \hline
  \noalign{\smallskip}
Secondary &  K3V--K5V   & $0.41\sin^{-3}{i}$ & $\sim 9.88$\,hr & -- & --\\
Primary   & White Dwarf &  $0.54\sin^{-3}{i}$ & $33.08$\,s & $5.64\times 10^{-14}$ & $74^{\degr} - 76^{\degr}$ \\
  \noalign{\smallskip}
  \hline
    \noalign{\smallskip}
\end{tabular}
\newline
$^*$~~For references see e.g. \citet{Ikhsanov-etal-2004}
\newline
$^{**}$~The angle between the rotational and magnetic axes
\end{table}


   \section{Diskless mass-transfer}

As first recognized by \citet{Wynn-etal-1997}, the observed Doppler H$\alpha$ tomogram
of AE~Aqr gives no evidence for a presence of an accretion disk in the system.
Simulating the mass-transfer in terms of drag interaction between the inhomogeneous
stream (a set of large diamagnetic blobs) and the magnetosphere (drag-driven propeller
model) they have shown that the material transferred from the normal companion through
the L1 point is streaming away from the system without forming a disk.

\begin{figure}
 \centerline{\resizebox{16cm}{!}{\includegraphics{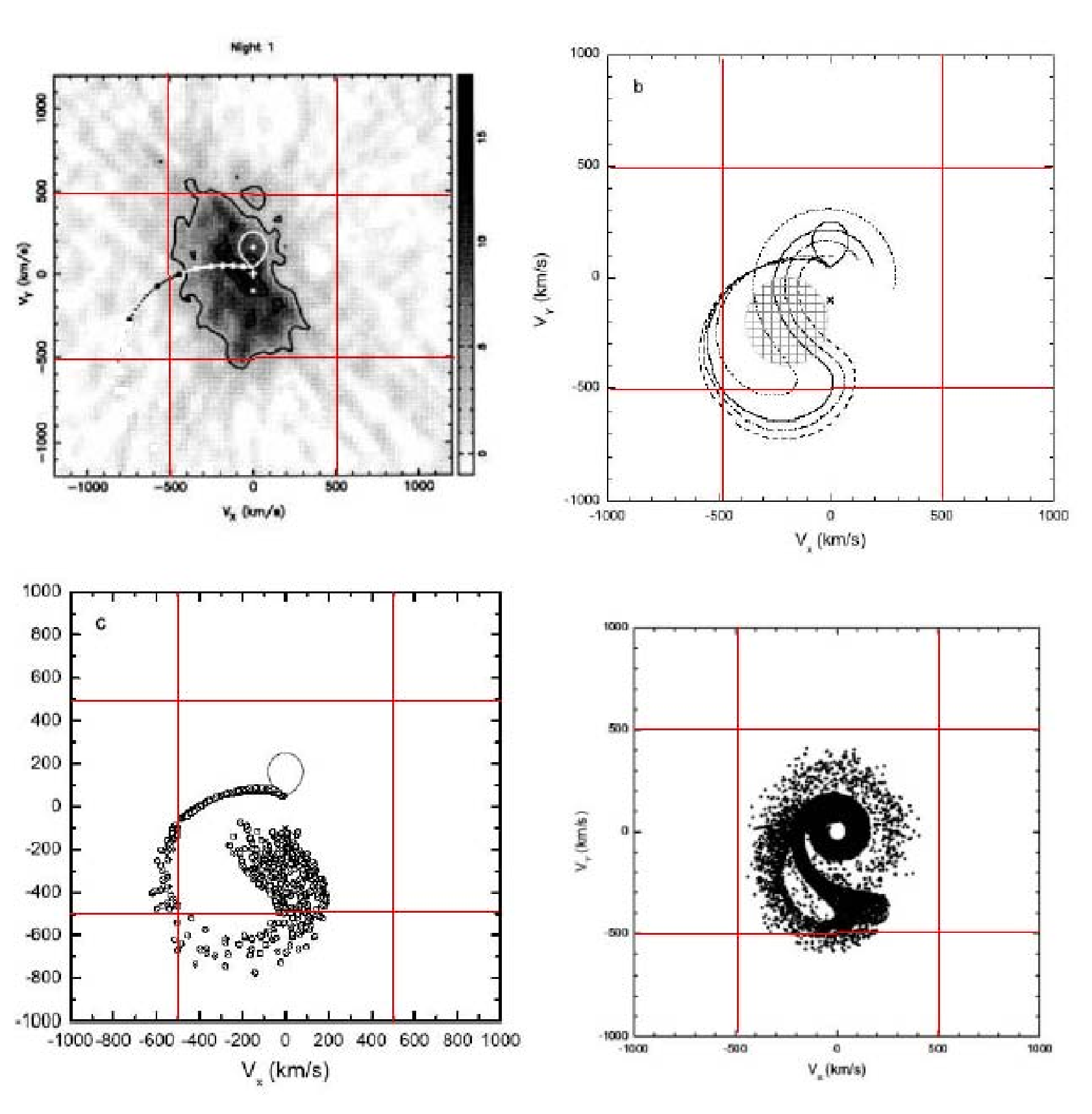}}}
\caption{Doppler H$\alpha$ tomogram of AE~Aquarii. Left-up: observed
\citep{Welsh-etal-1998}; Left-down: simulated under the assumption $\mu \sim
10^{32}\,{\rm G\,cm^3}$ \citep{Wynn-etal-1997,Ikhsanov-etal-2004}; Right-up: simulated
under the assumptions that $\mu \sim 10^{32}\,{\rm G\,cm^3}$ and the emission comes from
outside the system \citep{Welsh-etal-1998}; Right-down: simulated under the assumption
$\mu \simeq 1.5 \times 10^{34}\,{\rm G\,cm^3}$ \citet{Ikhsanov-etal-2004}. For further
discussion see \citet{Ikhsanov-etal-2004}.}
 \label{ikhsanov-f1}
\end{figure}

\citet{Ikhsanov-etal-2004} further elaborate on the latter picture concluding that the
structure of the simulated tomogram depends on the magnetic field strength of the white
dwarf. A best agreement between the observed and simulated tomograms has been found for
$\mu_{\rm wd} \simeq 1.5 \times 10^{34}\,{\rm G\,cm^3}$ (see Fig.\,\ref{ikhsanov-f1}).
Under these conditions the stream approaches the white dwarf to a distance
   \be\label{ralf}
r_0 \ga R_{\rm A} \simeq 3 \times 10^{10}\ \eta_{0.37}\ \mu_{34.2}^{4/7}\
\dmf_{17}^{-2/7}\ M_{0.9}^{-1/7}\ {\rm cm},
 \ee
where $R_{\rm A}$ is the Alfv\'en (magnetospheric) radius of the white dwarf with the
dipole magnetic moment, $\mu_{34.2}$, and mass, $M_{0.9}$, expressed in units of
$10^{34.2}\,{\rm G\,cm^3}$ and $0.9\,M_{\sun}$. The mass-transfer rate, $\dmf_{17}$, is
expressed in units of $10^{17}\,{\rm g\,s^{-1}}$, and $\eta_{0.37}=\eta/0.37$ is the
parameter accounting for the geometry of the accretion flow normalized following
\citet{Hameury-etal-1986}. Under the conditions of interest $R_{\rm A}$ exceeds the
circularization radius (see Tab.\,\ref{ikhsanov-t3}) and, therefore, prevents a
formation of an accretion disk in the system. The stream velocity at $R_{\rm A}$ is
limited to $\la 550\,{\rm rm\,s^{-1}}$. This is consistent with the upper limit to the
velocity of the flow in AE~Aqr derived from the observed Doppler H$\alpha$ tomogram
\citep{Welsh-etal-1998}. This provides us with a natural solution of the high-velocity
loop problem reported by \citet{Wynn-etal-1997}.

Thus, modeling of mass-transfer in AE~Aqr in terms of the drag-driven propeller model
suggests that the dipole magnetic moment of the white dwarf is $\mu \simeq 1.5 \times
10^{34}\,{\rm G\,cm^3}$.

%
 \begin{table}[t]
 \caption[]{Energy budget of AE~Aquarii$^{*}$}
 \label{ikhsanov-t2}
   \begin{tabular}{cccccccc}
         \noalign{\smallskip}
        \hline
       \noalign{\smallskip}
Component  & Balmer  & UV-Emission & H$\alpha$ &  X-rays &  Radio & $L_{\rm b}\,^{\dag}$ & $L_{\rm sd}\,^{\dag}$ \\
           & Continuum &  Lines &  & 0.1--20\,keV & 5--240\,MHz & & \\
         \noalign{\smallskip}
        \hline
       \noalign{\smallskip}
Quiescence$^{**}$  & $2.0 \times 10^{31}$ & $1.6 \times 10^{31}$ & $4.8 \times 10^{30}$ & $7.8 \times 10^{30}$ & $10^{28}$ &  $10^{33}$ & $6 \times 10^{33}$ \\
         \noalign{\smallskip}
        \hline
       \noalign{\smallskip}
Flares$^{**}$   & $8.4 \times 10^{31}$  & $4.1 \times 10^{31}$ & $1.4 \times 10^{31}$ & $1.7 \times 10^{31}$ & $2 \times 10^{29}$ & $10^{33}$ & $6 \times 10^{33}$  \\
         \noalign{\smallskip}
        \hline
       \noalign{\smallskip}
  \end{tabular}
  \newline
$^*$~~For references see e.g. \citet{Ikhsanov-etal-2004}, \citet{Ikhsanov-Biermann-2006}
\newline
$^{**}$~in erg\,s$^{-1}$
\newline
$^{\dag}$~$L_{\rm b}$ is the system bolometric luminosity and $L_{\rm sd}$ is the
spin-down power of the white dwarf
 \end{table}
%

 \section{Spin-down mechanism of the white dwarf}

   \subsection{Propeller spin-down}

The spin-down power by the magnetic propeller is limited to the magnetic flux transfer
rate through the region of interaction between the magnetic field and the blobs. This
condition can be expressed as
 \be\label{lkin}
L_{\rm prop} \la \frac{B^2(r_0)}{8 \pi}\ \sigma_{\rm eff}(r_0)\ N_{\rm b}\ t_{\rm
int}(r_0)\ |V_{\rm f}(r_0) - V{\rm b}(r_0)|_{\bot},
 \ee
where $B(r_0)=2 \mu_{\rm wd}/r_0^3$. The number of blobs approaching the white dwarf to
a distance $r_0$ in a unit time is limited to
 \be
 N_{\rm b} \la 3\ \times\ \dmf_{17}\ \rho_{-11}^{-1}\ {\it l}_9^{-3},
 \ee
where ${\it l}_9$ and $\rho_{-11}$ are the radius and density of the blobs expressed in
units of $10^9$\,cm and $10^{-11}\,{\rm g\,cm^{-3}}$ \citep[see][]{Wynn-etal-1997}.
Under the conditions of interest (i.e. $r_0 \gg R_{\rm cor}$) the field velocity,
$V_{\rm f}=\omega_{\rm s} r_0$, significantly exceeds the velocity of blobs, $V_{\rm
b}\la V_{\rm ff}=\left(2GM_{\rm wd}/r_0\right)^{1/2}$, and hence, $|V_{\rm f}(r_0) -
V{\rm b}(r_0)|_{\bot} \simeq \omega_{\rm s} r_0$ (the suffix $\bot$ denotes the velocity
component perpendicular to the field lines). The time of interaction between the blobs
and the magnetosphere at a distance $r_0$ is of the order of free-fall time, $t_{\rm
int} \simeq \left(r_0^3/2GM_{\rm wd}\right)^{1/2}$. Finally, the effective cross-section
of interaction between the blob and magnetic field can be expressed as
 \be\label{sigma}
\sigma_{\rm eff} \simeq 2 \pi {\it l}_{\rm b} \Delta r \sim 2 \pi {\it l}_{\rm b}
\left(D_{\rm eff}\ t_{\rm int}\right)^{1/2},
  \ee
where $\Delta r$ is a scale of the magnetic field diffusion into the blob on a time
scale $t_{\rm int}$. In the case of Bohm diffusion \citep[i.e. $D_{\rm eff} =
\alpha_{\rm B} \dfrac{ck_{\rm B}T_{\rm i}}{16 e B}$, see,
e.g.][]{Ikhsanov-Pustilnik-1996} one finds (combining Eqs.~\ref{lkin} - \ref{sigma})
 \be
L_{\rm prop} \la 10^{32}\,{\rm erg\,s^{-1}}\ \alpha_{0.1}\ N_{\rm b}\ \omega_{0.2}\ {\it
l}_9\ \mu_{34}^{3/2}\ M_{0.8}^{-3/7}\ T_8^{1/2}\ \left(\frac{r_0}{3 \times 10^{10}\,{\rm
cm}}\right)^{-5/4} ~ \simeq 2 \times 10^{-2}\ L_{\rm sd},
 \ee
where $\alpha_{0.1} = \alpha_{\rm B}/0.1$ is the diffusion efficency, which is
normalized following the resulats of measurements of solar wind penetrating into the
magnetosphere of the Earth \citep{Gosling-etal-1991},  $\omega_{0.2}=\omega_{\rm s}/0.2$
and $T_8$ is the temperature of outer layers of the blobs expressed in units of
$10^8$\,K.

Thus, the efficiency of the propeller action by the white dwarf constitutes only a few
per cents of the observed spin-down power. This indicates that the observed braking of
the white dwarf is governed by a different mechanism.

%
\begin{table}[t]
 \caption[]{Basic scales of AE~Aquarii}
 \label{ikhsanov-t3}
 \begin{tabular}{ccccccc}
    \noalign{\smallskip}
     \hline
      \noalign{\smallskip}
 Radius$^*$   &  $R_{\rm wd}$ &  $R_{\rm cor}$ & $R_{\rm circ}$ & $R_{\rm L1}$ &  $R_{\rm lc}$ & $a$ \\
    \noalign{\smallskip}
     \hline
      \noalign{\bigskip}
Value (cm) & ~ $(6-7) \times 10^8$ & $1.5 \times 10^9$ &  $(1.8-2) \times
10^{10}$  &~ $10^{11}$  ~ & $1.6 \times 10^{11}$ &  $(1.7-1.9) \times 10^{11}$ \\
      \noalign{\smallskip}
     \hline
      \noalign{\bigskip}
  \end{tabular}
  \newline
$^*$~$R_{\rm wd}$ is the white dwarf radius; $R_{\rm cor}$ is the corotational radius;
$R_{\rm circ}$ is the circularization radius; $R_{\rm L1}$ is the distance from the
white dwarf to the L1 point; $R_{\rm lc}$ is the radius of the light cylinder; $a$ is
the orbital separation of the system components.
  \end{table}



 \subsection{Pulsar-like spin-down}

The only astrophysical objects whose spin-down power significantly exceeds their
bolometric luminosity are the spin-powered pulsars. The spin-down power of these objects
is released mainly in a form of magneto-dipole waves and accelerated particles. As
recently shown by \citet{Ikhsanov-Biermann-2006}, the spin-down mechanism developed for
these objects is perfectly applicable to the case of the white dwarf in AE~Aqr. In
particular, the observed braking of the white dwarf can be explained in terms of the
pulsar-like spin-down mechanism provided that its dipole magnetic moment is
   \be\label{magmom}
\mu \simeq 1.4 \times 10^{34}\ \left(\frac{P_{\rm s}}{33\,{\rm s}}\right)^{2}
\left(\frac{L_{\rm sd}}{6 \times 10^{33}\,{\rm erg\,s^{-1}}}\right)^{1/2} {\rm
G\,cm^{3}}.
   \ee
This implies that the strength of the surface magnetic field of the white dwarf in the
magnetic pole regions is
 \be\label{B0}
 B_0 = \frac{2 \mu}{R_{\rm wd}^3} \simeq 100\ \left(\frac{R_{\rm wd}}{7 \times 10^8\,{\rm cm}}\right)^{-3}\
 \left[\frac{\mu}{1.4 \times 10^{34}\,{\rm G~cm^3}}\right]\ {\rm MG},
 \ee
and, correspondingly, the surface field strength at its magnetic equator is $B_0/2 =
50$\,MG.

As seen from Eq.~(\ref{magmom}), the value of $\mu$ required for the interpretation of
the rapid braking of the white dwarf in terms of the pulsar-like spin-down mechanism is
in excellent agreement with that derived by \citet{Ikhsanov-etal-2004} from the
simulation of the Doppler H$\alpha$ tomogram within the drag-driven propeller approach.
Thus, analysis of both the mass-transfer and the braking of the white dwarf suggests the
same value of the dipole magnetic moment of the white dwarf, namely, $\mu \simeq 1.5
\times 10^{34}\,{\rm G\,cm^3}$.

 \section{Circularly polarized optical emission}

As first shown by \citet{Cropper-1986}, the optical emission of AE~Aqr is circularly
polarized at a level of $p = 0.05\% \pm 0.02\%$. An attempt to explain this result in
terms of the cyclotron emission from the base of the accretion column has led
\citet{Bastian-etal-1988} to a serious problem. Namely, they have found that
observations reported by \citet{Cropper-1986} can be interpreted in terms of the model
of \citet{Chanmugam-Frank-1987} only if the surface field of the white dwarf exceeds
$10^6$\,G. However, the magnetospheric radius of the white dwarf (see Eq.~\ref{ralf})
under these conditions is significantly larger than its corotational radius (see
Tab.\,\ref{ikhsanov-t3}) and hence, an accretion of material onto the surface of the
white dwarf is impossible (the star turns out to be in the centrifugal inhibition
state). But if this is really so, the model of \citet{Chanmugam-Frank-1987} is not
applicable (because of a lack of the accretion column) and cannot be used for
determination of the surface field of the white dwarf.

In order to solve this paradox \citet{Beskrovnaya-etal-1996} have performed an
independent measurement of the degree of circularly polarized emission of the system.
Their result, $p = 0.06\% \pm 0.01\%$, is in excellent agreement with the result of
\citet{Cropper-1986}. Analyzing these observations \citet{Ikhsanov-etal-2002} have shown
that the hot polar caps at the surface of the white dwarf, which were recognized through
observations of the system with the Hubble space telescope \citep{Eracleous-etal-1994},
cannot be a source of the observed circularly polarized emission. Otherwise, the
intrinsic polarization of the optical emission in these regions turns out to be in
excess of 100\%. It has also been shown that the polarized emission cannot be
interpreted in terms of the linear and quadratic Zeeman effect. The results of numerical
simulations reported by \citet{Ikhsanov-etal-2002} show that the degree of polarization
in the case of $B = 1$\,MG is close to zero and for the case of $B = 50$\,MG is limited
to $\la 0.015\%$, which is a factor of 4 smaller than the observed value.

Thus, a question about the mechanism responsible for the observed circularly polarized
optical emission remains so far open. It is, however, absolutely clear that the
evaluation of the magnetic field reported by \citet{Bastian-etal-1988} has been made
under assumptions which are not valid in the case of AE~Aqr and, therefore, does not
reflect the system properties.

   \section{Pulsing X-ray emission}

As recently reported by \citet{Terada-etal-2008}, properties of pulsing X-ray emission
of the system observed with SUZAKU at 10-30\,keV suggest a non-thermal nature of this
radiation. The authors of this discovery have associated this emission with radiative
losses of electrons accelerated in the magnetosphere of the fast rotating, strongly
magnetized white dwarf. The luminosity of the non-thermal pulsing X-ray source has been
evaluated as $\sim (0.5-2.3) \times 10^{30}\,{\rm erg\,s^{-1}}$. It has been noted,
however, that the hard X-ray pulsations have a duty ratio of only 0.1, which may reflect
the fact that the radiation is anisotropic and highly beamed. This indicates that the
source luminosity can be limited to\footnote{This value of $L_{\rm min}$ is smaller that
that presented by \citet{Terada-etal-2008} by a factor of 100. It appears that
\citet{Terada-etal-2008} have mistakenly evaluated the luminosity of the beamed source
by multiplying the luminosity of the isotropic source by ($4 \pi/\gamma_{\rm col}$)
instead of dividing it by the same value, where $\gamma_{\rm col}$ is the opening body
angle of the beam.} $L_{\rm x-p} \ga L_{\rm min} \simeq 5 \times 10^{28}\,{\rm
erg\,s^{-1}}$.

It is widely believed that acceleration of particles by the white dwarf in AE~Aqr
represents nothing unusual since the electric potential in the magnetosphere of this
star ($V_{\rm e} \sim 2\pi R_{\rm wd}^2 B_{\rm wd}/P_{\rm s}$) has a huge value ($\sim
10^{14}-10^{16}$\,V). However, for an acceleration of particles in this potential to be
effective the number density of material in the magnetosphere should not exceed the
Goldreich-Julian density
 \be\label{ngj}
n_{\rm GJ} = \frac{(\vec{\Omega} \cdot \vec{B})}{2 \pi c e} \simeq 5 \times 10^4\
\left(\frac{P_{\rm s}}{33\,{\rm s}}\right)^{-1}\ \left(\frac{B}{10^8\,{\rm G}}\right)\
{\rm cm^{-3}}.
 \ee
Otherwise, the electric field responsible for particle acceleration would be screened by
the magnetospheric plasma. As recently shown by \citet{Ikhsanov-Biermann-2006}, the
kinetic luminosity of the beam of relativistic particles under these conditions is
limited to
 \be\label{dedt}
\frac{d \mathcal{E}_{\rm p}}{dt} \la  e\ \varphi(l_0)\ \dot{N}\ \simeq \ 5 \times
10^{29}\,{\rm erg\,s^{-1}} \left(\frac{B_{\rm wd}}{10^8\,{\rm G}}\right)^2,
 \ee
where $e$ is the electron electric charge and $l_0=(R_{\rm wd}+s)$ is a distance from
the surface of the white dwarf to the region of generation of the X-rays.
 \be\label{phi}
 \varphi_{\rm as}(l_0) = \int_{\rm R_{\rm wd}}^{l_0} E_{\parallel}\ ds\ \simeq\
 2 \sqrt{2}\ E_{\rm AS}\ R_{\rm wd} \left[\left(\frac{l_0}{R_{\rm
wd}}\right)^{1/2} - 1\right]
 \ee
is the electric potential generated in the polar cap regions of a fast rotating
magnetized star surrounded by a vacuum \citep{Arons-Scharlemann-1979}. $E_{\parallel}
\equiv(\vec{E} \cdot \vec{B})/|\vec{B}|$ is the component of the electric field along
the magnetic field $\vec{B}$, and
 \be\label{eas}
E_{\rm AS} = \frac{1}{8\sqrt{3}} \left(\frac{\Omega R_{\rm  wd}}{c}\right)^{5/2}
B(R_{\rm wd}).
 \ee
The flux of relativistic particles from the polar caps of the white dwarf is
 \be\label{ndot}
 \dot{N} = \pi (\Delta  R_{\rm p})^2 n_{\rm GJ}(R_{\rm wd}) c,
 \ee
where
 \be
 \Delta R_{\rm p} \simeq \left(\frac{\omega R_{\rm wd}}{c}\right)^{1/2}
R_{\rm wd} \simeq 4 \times 10^7\ \left(\frac{R_{\rm wd}}{7 \times 10^8\,{\rm
cm}}\right)^{3/2}\ \left(\frac{P_{\rm s}}{33\,{\rm s}}\right)^{-1/2}\ {\rm cm}
 \ee
is the radius of the polar caps.

As follows from Eq.~(\ref{dedt}), the observed luminosity of the hard X-ray pulsing
component can be explained in terms of the pulsar-like acceleration mechanism only if
the surface field of the white dwarf satisfies the condition
 \be
B(R_{\rm wd})\ \ga\ 3 \times 10^7\,{\rm G}\ \times\ \eta^{-1}\ \left(\frac{L_{\rm
min}}{5 \times 10^{28}\,{\rm erg\,s^{-1}}}\right)^{1/2},
 \ee
where $\eta <1$ is the efficiency of conversion of the energy of accelerated particles
into the X-rays. This indicates that the dipole magnetic moment of the white dwarf is
$\ga 10^{34}\,{\rm G\,cm^3}$, i.e. close to the value derived from the simulation of the
Doppler H$\alpha$ tomogram of the system and from the modeling of the braking of the
white dwarf in terms of the pulsar-like spin-down.

  \section{Discussion and conclusions}

Simulations of Doppler H$\alpha$ tomogram, analysis of the rapid braking of the white
dwarf, and the modeling of particle acceleration in AE~Aqr suggest that the dipole
magnetic moment of the white dwarf in this system is $\mu \simeq 1.5 \times
10^{34}\,{\rm G\,cm^3}$. Under these conditions the spin-down of the white dwarf is
governed by the pulsar-like mechanism, i.e. its spin-down power is released
predominantly in a form of the magneto-dipole waves and accelerated particles and
significantly exceeds the bolometric luminosity of the system. This makes the degenerate
companion of AE~Aqr the first white dwarf in the family of spin-powered pulsars.

The above evaluation of the magnetic field of the white dwarf does not contradict
observations of the circularly polarized optical emission as well as all presently
established properties of the system. Moreover, the pulsar-like approach appears to be
an effective tool in explanation of the origin of hot polar caps at the white dwarf
surface \citep[in terms of the dissipation of the backflowing current in the
magnetosphere, see, e.g.][]{Ikhsanov-etal-2004, Ikhsanov-Biermann-2006} and properties
of the system derived from observations in high-energy part of the spectrum.

At the same time, the above conclusion rises a problem about the history of the system.
The age of the white dwarf evaluated from its surface temperature  $(1-1.6) \times
10^4$\,K \citep{Eracleous-etal-1994} is limited to $\ga 10^8$\,yr \citep[see,
e.g.][]{Schoenberner-etal-2000}, while the spin-down time scale is only $P/\dot{P} \sim
10^7$\,yr. This indicates that the system history contains an accretion-driven spin-up
epoch. However, the spin period to which a white dwarf with $\mu \sim 10^{34}\,{\rm
G\,s^{-1}}$ could be spun-up by a disk accretion is substantially larger than the
currently observed 33\,s period \citep[for discussion see][]{Ikhsanov-1999}. It,
therefore, appears, that the magnetic field of the white dwarf has been amplified during
a previous epoch. As shown by \citet{Ikhsanov-1999}, this could occur at the end of the
accretion-driven spin-up phase due to gravitational waves emission instability of the
degenerate core of the star. In this light, AE~Aqr can be considered as a precursor of a
polar. The detailed study of the corresponding evolutionary track isnow work in
progress.

\vspace{1cm}

Nazar Ikhsanov acknowledge the support of the European Commission under the Marie Curie
Incoming Fellowship Program. This work was partly supported by Russian Foundation of
Basic Research under the grant number 07-02-00535a.

\end{document}